\documentclass{emulateapj}

\usepackage{epsfig}

\newcommand{\etal}{\mbox{et al.}}
\newcommand{\ergcms}{erg cm$^{-2}$ s$^{-1}$}
\newcommand{\ergs}{erg s$^{-1}$}
\newcommand{\phcms}{ph cm$^{-2}$ s$^{-1}$}

\newcommand{\msun}{$M_{\odot}$}
\newcommand{\pc}{\,{\rm pc}}
\newcommand{\yr}{\,{\rm yr}}

\newcommand{\emsun}{\,M_\odot}
\newcommand{\au}{\,{\rm AU}}

\newcommand{\chandra}{{\it Chandra}}

\newcommand{\bepposax}{{\it BeppoSAX}}
\newcommand{\xmm}{{\it XMM-Newton}}

\newcommand{\rxte}{{\it RXTE}}

\newcommand{\sgrastar}{\mbox{Sgr A$^*$}}

\makeatletter
\newenvironment{inlinefigure}{%
\def\@captype{figure}%
\noindent\begin{minipage}{0.999\linewidth}\begin{center}}
{\end{center}\end{minipage}\smallskip}

\newcommand\newtablebreak{\cr\ptable@@split}
\makeatother

\shortauthors{Muno \etal}
\shorttitle{X-ray Transients at the Galactic Center}

\submitted{} 
\begin{document}
\title{An Overabundance of Transient X-ray Binaries within 1 pc of 
the Galactic Center}
\author{M. P. Muno,\altaffilmark{1,2} E. Pfahl,\altaffilmark{3}
F. K. Baganoff,\altaffilmark{4} 
W. N. Brandt, \altaffilmark{5}
A. Ghez,\altaffilmark{1}
J. Lu,\altaffilmark{1}
and M. R. Morris\altaffilmark{1} 
}

\altaffiltext{1}{Department of Physics and Astronomy, University of California,
Los Angeles, CA 90095; mmuno@astro.ucla.edu}
\altaffiltext{2}{Hubble Fellow}
\altaffiltext{3}{Chandra Fellow, Department of Astronomy, University of 
Virginia, Charlottesville, VA 22904}
\altaffiltext{4}{Kavli Institute for Astrophysics and Space Research,
Massachusetts Institute of Technology, Cambridge, MA 02139}
\altaffiltext{5}{Department of Astronomy and Astrophysics, 
The Pennsylvania State University, University Park, PA 16802}

\begin{abstract}
During five years of Chandra observations, we have identified 
seven X-ray transients located within 23 pc of \sgrastar.
These sources each vary in luminosity by more than a factor of 10, and 
have peak X-ray luminosities greater than $5 \times 10^{33}$ \ergs, 
which strongly suggests that they are accreting black holes or 
neutron stars. The peak luminosities of the transients are intermediate 
between those typically considered 
outburst and quiescence for X-ray binaries. Remarkably four of these 
transients lie within only 1 pc of \sgrastar. This implies that, compared 
to the numbers of similar systems located between 1 and 23 pc, 
transients are over-abundant by a factor of $\approx 20$ per unit stellar 
mass within 1 pc of \sgrastar.
It is likely that the excess transient X-ray sources are
low-mass X-ray binaries that were produced, as in the cores of globular 
clusters, by three-body
interactions between binary star systems and either black holes or 
neutron stars that have been concentrated in the central parsec through 
dynamical friction. Alternatively, they could be high-mass X-ray
binaries that formed among the young stars that are present in the 
central parsec.
\end{abstract}


\section{Introduction}

\chandra\ has observed the inner 20 pc of the Galaxy repeatedly for 
the past five years, primarily to monitor the super-massive black hole
\sgrastar\ \citep{bag03}. However, the combination of the repeated 
observations, the 
exquisite sensitivity of \chandra, and the high stellar density in the field
also is ideal for detecting and studying transient X-ray sources. 
The total mass of stars in a pencil beam enclosing the inner 20~pc of the 
Galaxy is $\sim 10^{8}$~\msun, or $\sim 0.1$\% of the Galactic stellar 
mass. Therefore, these observations sample a huge number of stars, and are 
likely to reveal rare X-ray sources such as accreting black holes and neutron 
stars. Moreover, while traditional all-sky 
X-ray monitors detect sources as faint as $\simeq 10^{36}$~\ergs\ 
near the Galactic center \citep[e.g.,][]{lev96,jag97},
even short \chandra\ observations are sensitive to sources three orders of 
magnitude fainter. 
As a result, these observations can be used to study a little-observed 
regime of accretion for X-ray binaries between outburst 
($L_{\rm X} > 10^{36}$~\ergs) and quiescence ($L_{\rm X} < 10^{33}$~\ergs;
see Sakano \etal\ 2004).

The distribution of X-ray binaries in the Galactic 
center could also yield insight into the interaction between the
central super-massive black hole and the surrounding dense concentration of
stars. First, tens of thousands of stellar-mass black holes and
neutron stars are likely to have settled dynamically into the central 
parsec of the Galaxy over the past few Gyr \citep{mor93,lee95,meg00}. 
Three-body interactions 
between this concentration of compact objects and background binary stars can 
form X-ray binaries (E. Pfahl \& A. Loeb, in prep.), in a similar 
manner to that by which 
low-mass X-ray binaries (LMXBs) form in globular 
clusters \citep[e.g.,][]{rpr00}.
Second, several dozen stars only 7~Myr old have been discovered $< 1$~pc 
from \sgrastar\ \citep[e.g.,][]{kra95}, and high-mass X-ray 
binaries (HMXBs) may have formed among this young stellar population. 
Therefore, we have searched the \chandra\ observations of \sgrastar\ for 
transient X-ray sources
that are likely to be accreting black holes and neutron stars. 

\section{Observations and Results} 

The region within 10\arcmin\ of \sgrastar\ has been observed sixteen
times between 1999--2004 with the Advanced CCD Imaging Spectrometer 
imaging array (ACIS-I) aboard the {\it Chandra X-ray Observatory} 
\citep{wei02}. The first twelve observations are listed 
in Table~2 of \citet{mun03a}. The remaining four were taken on 2003 June 19 
for 25 ks (sequence 3549), 2004 July 5--7 for 99 ks (sequences 4683 and 4684), 
and 2004 August 28 for 5 ks (sequence 5630).
We reduced the data and searched for X-ray sources using the techniques 
explained in \citet{mun03a}.

We used three criteria to select transient X-ray sources that were most likely 
to be accreting black holes and neutron stars near the Galactic center
\citep[compare, e.g.,][]{cam98}. 
First, in order to avoid including foreground sources, we selected those
sources with X-ray spectra that were absorbed by a column of gas and dust
similar to or greater than that toward \sgrastar, which is equivalent to 
$5\times10^{22}$ cm$^{-2}$ of hydrogen \citep{bag03}. This was implemented
by requiring that the soft color, defined as in \citet{mun03a}
using the 0.5--2.0 and 2.0--3.3 keV energy bands, was larger than --0.175. 

Second, we selected only those sources with peak luminosities  
$\ga 5\times10^{33}$~\ergs, because the fainter sources are 
most likely cataclysmic variables \citep[][]{ver97}. 
For a distance of 8 kpc \citep{rei99} and a 
$\Gamma = 1.5$ power law spectrum, our 
threshold corresponds to fluxes greater than $2 \times 10^{-5}$ 
photon cm$^{-2}$ s$^{-1}$ (2--8 keV). The 
flux limit at which a source 
can be detected 
at the 3-$\sigma$
 level within 10\arcmin\ of the aim-point of a 5~ks 
observation is 

\begin{inlinefigure}
\centerline{\psfig{file=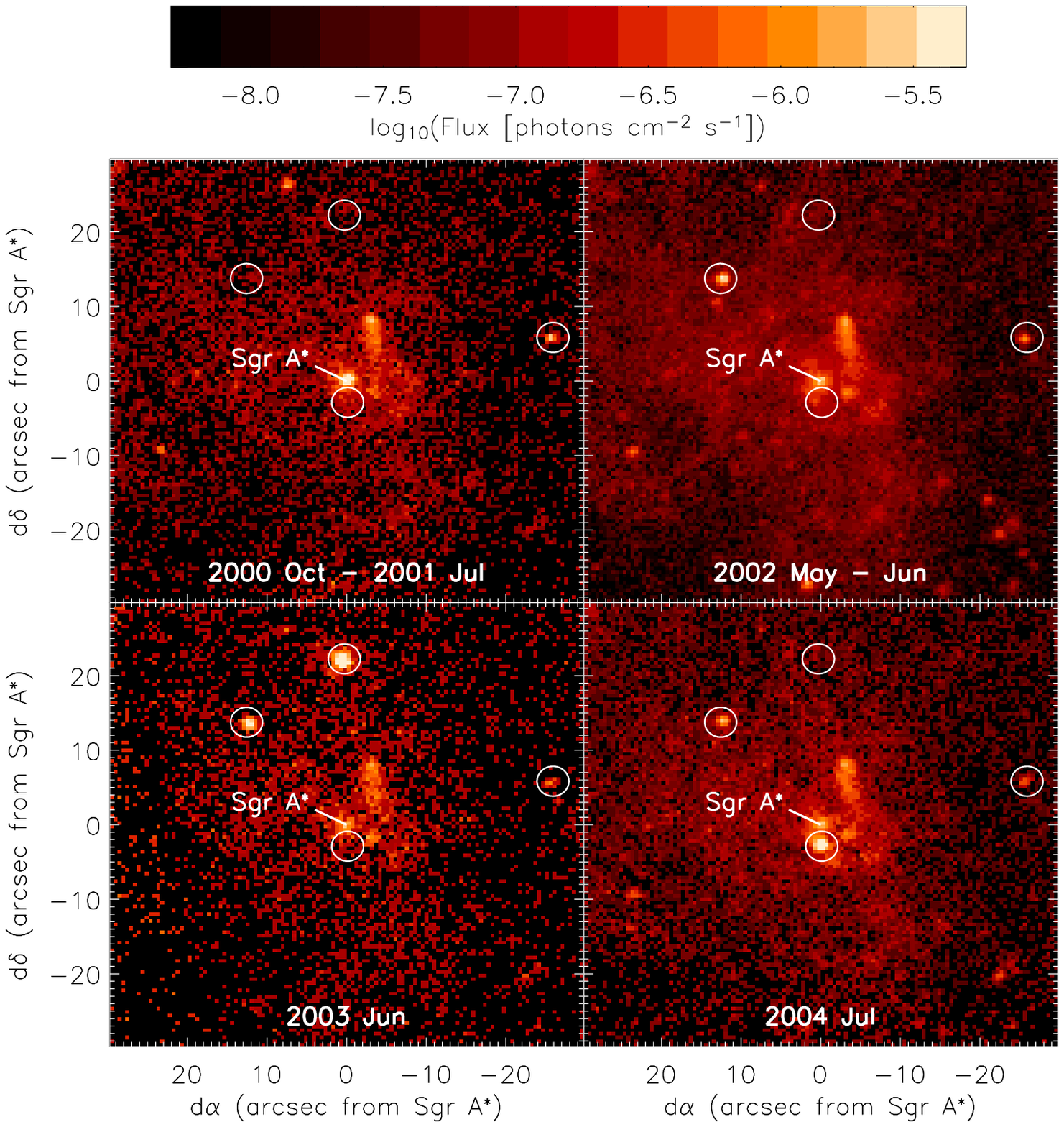,width=\linewidth}}
\caption{
Chandra images of the 1\arcmin\ by 1\arcmin\ field around \sgrastar, taken
at several epochs. Four of the transient X-ray sources are present in this 
image, and their locations are indicated schematically by circles. 
}
\label{fig:ximg}
\end{inlinefigure}

\noindent
$ 5\times10^{-6}$ photon cm$^{-2}$ s$^{-1}$ 
\citep[Fig. 7 in][]{mun03a}, so we are uniformly sensitive to 
sources this bright located within 23 pc in projection from \sgrastar.

Finally, in order to select accreting sources with large transient outbursts,
we required the flux from the source to vary by at least a factor of 10. This 
eliminates bright, steady X-ray sources such as young pulsars 
and Wolf-Rayet and O stars in binaries.\footnote{
One known 
cataclysmic variable would pass our selection criteria --- 
GK Per exhibits X-ray outbursts lasting 45 days that brighten by a factor 
of $\approx 10$ to peak luminosities of $1\times10^{34}$~\ergs\ 
\citep[e.g.,][]{krw79}. None of the outbursts from our sources appears 
similar to that of GK Per, as they each either last longer than a year, or 
brighten by more than a factor of 100.
}

Seven sources satisfied our selection criteria, and are listed in 
Table~\ref{tab:sources}. Images displaying the four 
transients closest to \sgrastar\ are displayed in Figure~\ref{fig:ximg}.

\subsection{Properties of the Transients}

We have examined the light curves of each of the transients
using the techniques described in \citet{mun04b}.
We display the flux history of the sources in 
Figure~\ref{fig:lcurves}. We have omitted CXOGC J174502.3--285450 
(GRS 1741.9--2853) because this source is described in Muno, 
Baganoff, \& Arabadjis (2003)\nocite{mun03b}.
In that paper, we describe a thermonuclear X-ray burst 
lasting $\approx 50$~s that identifies the source as a neutron-star 
LMXB. No similar, short bursts were observed from the other transients.

We searched for periodic variability as described in \citet{mun03c}. 
We find no evidence for signals with periods between 10--5000~s, with 
typical upper limits on their rms amplitudes ranging from 6\% for 
CXOGC J174554.3--285454\footnote{The upper limit on the amplitude of 
the 172~s signal reported by \citet{por04} is 4\% rms.} 
to 26\% for CXOGC J174538.0--290022. Most signals at longer periods are
produced by slow, random variations in the intensities of each source. 
However, a signal and its 
harmonic are detected with a period of $7.9$~h in the 2004 July 
observations of CXOGC~J174540.0--290031. The modulation is highly significant
(chance probability $< 10^{-10}$), and resembles  
eclipses that recur at the orbital period. Unfortunately, the
relevant individual observations were less than 14~h long, so this 
result is not
yet secure.

\begin{deluxetable}{lccc}
\tablecolumns{4}
\tablewidth{0pc}
\tablecaption{Transient X-ray Sources in the inner 10\arcmin\ of the 
Galaxy\label{tab:sources}}
\tablehead{
\colhead{Source} & 
\colhead{Offset} & 
\colhead{Min $L_{\rm X}$} & \colhead{Max $L_{\rm X}$} \\
\colhead{(CXOGC J)} & 
(arcmin) & 
\multicolumn{2}{c}{2--8 keV (\ergs)}
} 
\startdata
174502.3--285450 & 
9.98 & $<7\times10^{31}$ & $1.5\times10^{36}$ \\
174535.5--290124 & 
1.35 & $<9\times10^{30}$ & $3.3\times10^{34}$ \\
174538.0--290022 & 
0.44 & \phantom{$<$}$1.2\times10^{33}$ & $2.6\times10^{34}$ \\
174540.0--290005 & 
0.37 & $<4\times10^{31}$ & $3.4\times10^{34}$\\
174540.0--290031 & 
0.05 & $<2\times10^{31}$ & $8.5\times10^{34}$\\
174541.0--290014 & 
0.31 & $<8\times10^{31}$ & $4.8\times10^{33}$ \\
174554.3--285454 & 
6.38 & $<2\times10^{31}$ & $6.2\times10^{34}$
\enddata
\tablecomments{
We report 
$L_{\rm X}$ for the 2--8 keV band, because the inferred flux at lower 
energies depends heavily upon the assumed absorption column toward each 
source. The bolometric luminosities should be a factor
of 3--10 larger. Three sources are 
not listed in the catalog of \citep{mun03a}, so we give their positions
here: 
CXOGC J174554.3--285454: $\alpha = 266.4769$, $\delta = -28.9153$ 
(XMMU J174554.4--285456 in Porquet \etal\ 2004);
CXOGC J174540.0--290005: $\alpha = 266.4170$, $\delta = -29.0016$;
and CXOGC J174540.0--290031: $\alpha = 266.4168$, $\delta = -29.0086$.
}
\end{deluxetable}

%

Spectral results for most of the transients can be found in 
\citet{mun03b}, \citet{mun04b}, and \citet{por04}.
The spectra are diverse, but are consistent with the 
variety observed from known X-ray binaries. Otherwise, the spectra
do not yield any clues as to the natures of the 
mass donors or compact objects.


To constrain better the natures of the mass donors, we have searched for 
infrared counterparts to the four transient sources within 1\arcmin\ of 
\sgrastar\ using archival NICMOS observations and engineering observations 
taken with the Keck AO laser guide star system. 
We find that three of the 
four sources have potential infrared counterparts with $K \approx 15-17$, 
although the high density of stars makes it likely that these are random
associations. We find that there is a $\approx 50$\% chance that three 
of four randomly-placed, 0\farcs3 circles would contain a star with $K<16$. No 
observations of comparable depth are available for the sources located 
1--10\arcmin\ from \sgrastar.

Finally, VLA observations have revealed a radio outburst coincident with
the appearance in X-rays of CXOGC~J174540.0--290031, with a peak intensity
of 100 
mJy at 1 GHz (G. Bowers, F. Yusef-Zadeh, \& D. Roberts, in prep.). The 
large radio outburst suggests that this source is more likely to contain 
a black hole than a neutron star 
primary \citep[e.g.,][]{fk01}.

\subsection{Spatial Distribution of the Transients}

Remarkably, four of the seven transient X-ray sources are located 
within 0\farcm45 of \sgrastar, which corresponds to a projected distance 
of $<1$~pc. This concentration of transients near \sgrastar\
is significant even when one takes into account the $R^{-2}$ increase
in stellar density approaching

\begin{inlinefigure}
\centerline{\psfig{file=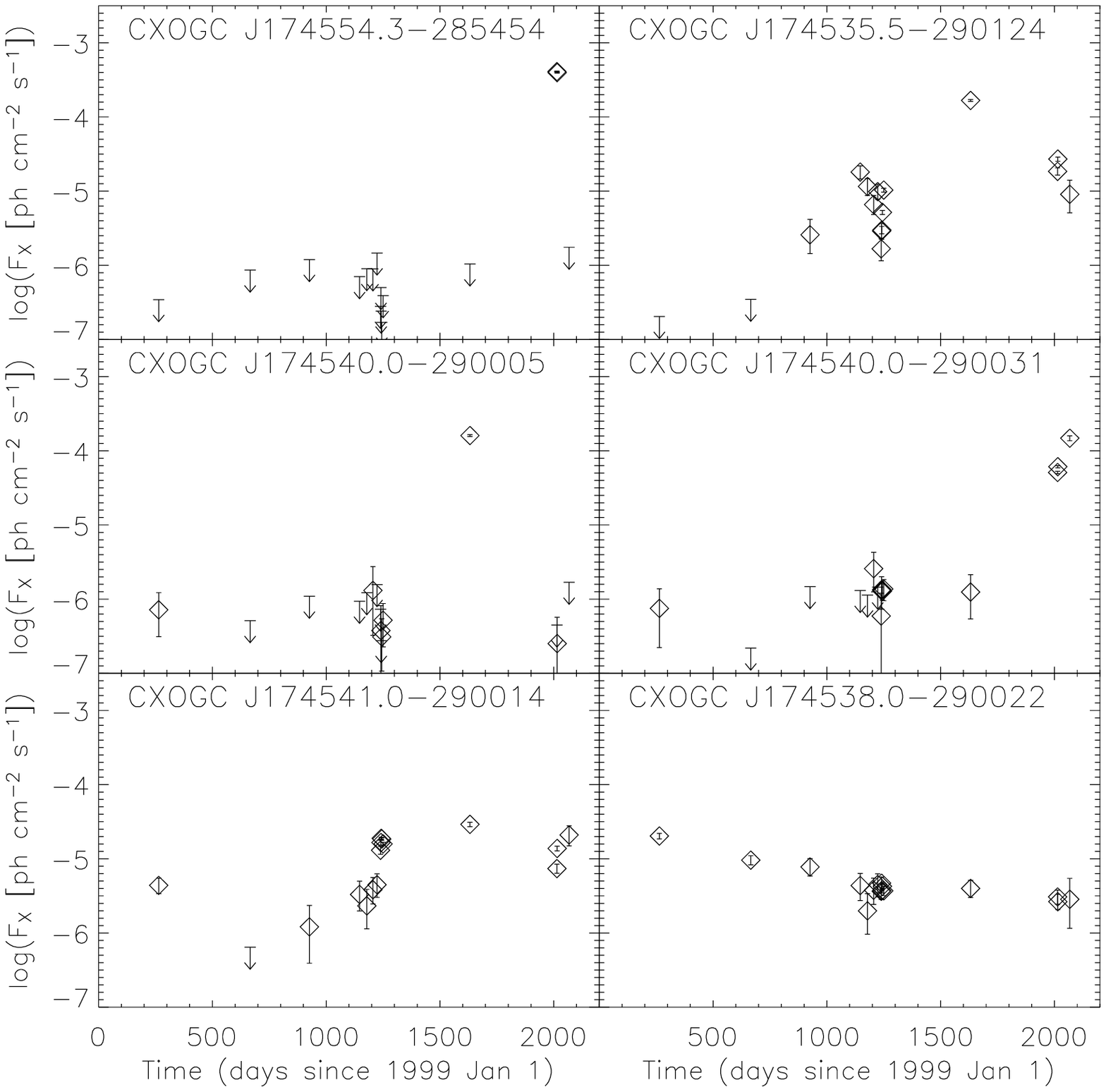,width=0.9\linewidth}}
\caption{
The long-term flux history of the six transient sources not displayed 
in \citet{mun03b}.
}
\label{fig:lcurves}
\end{inlinefigure}

\noindent
\sgrastar. 
Using 
the models of \citet{lzm02}, we find that the stellar mass within 
a sphere of 1~pc radius centered on \sgrastar\ is 
$2\times10^6$ \msun, whereas the mass in the sphere between 1 and 
23~pc (10\arcmin) is $3\times10^7$ \msun. This implies that there is 
a factor of 20 overabundance of bright transients within 1 pc of 
\sgrastar. 

We have computed the significance of this excess using the Fisher 
Exact Probability test. As a null hypothesis, we take the observed 
positions of Galactic center X-ray sources \citep{mun03a}, 
and assume that the 7 transients should be randomly distributed 
among these. For the combined set of observations, above the 
completeness limit ($F_{\rm X} = 5\times10^{-7}$ \phcms) we find 
24 sources within 1 pc of \sgrastar, and 464 sources between 
1 and 23 pc. The probability 
that $\ge 4$ of the 7 transients would randomly lie within 1 pc of 
\sgrastar\ is $2\times10^{-4}$.



\section{Discussion}

These transients are unusual because they were observed
at luminosities intermediate between those considered outburst and 
quiescence for X-ray binaries \citep{cam98,mer00}. 
Indeed, the class of faint X-ray transients discussed by 
\citet{king00} have peak luminosities that are 
$\ga 10$ times larger than those in our sample.
X-ray transients are typically identified using all-sky monitors
such as those on \rxte\ and \bepposax, which detect sources with fluxes 
$\ga 10^{-10}$~\ergcms\ \citep{lev96,jag97}. As a result, all-sky
monitors can detect transients similar to those in Table~\ref{tab:sources} 
out to a distance of only 600~pc. Within this distance, there are few known 
accreting black holes or neutron stars. 
Therefore, continued observations of 
the Galactic center will yield insight into the duty cycles of 
X-ray binaries accreting at low average rates of $\la 10^{-12}$
\msun\ yr$^{-1}$ \citep[cf.][]{sak04}.


Unfortunately, the relative lack of observations of X-ray binaries with 
$L_{\rm X} = 10^{33} - 10^{35}$~\ergs\ makes it difficult to establish
conclusively the natures of these transients. The fact that they are 
transient suggests that the systems are either LMXBs with unstable accretion
disks, or HMXBs in which either the mass-loss rate from the secondary is highly
variable or the compact primary has an eccentric orbit about
its companion. The lack of counterparts with $K < 15$ suggests that
the transients are either HMXBs with companions fainter than a B2IV star 
\citep{weg94,weg00}, or LMXBs. Finally, the 
candidate 7.9~h period in CXOGC~J174540.0--290031 is too short to accommodate
anything other than an LMXB,\footnote{The radius of a Roche-lobe-
filling mass donor would be $\approx 0.8 R_\odot$, 
which rules out a high-mass star (see Frank, King, \& Raine 1995, eq. 4.10).} 
while its bright radio outburst suggests that it 
contains a black hole \citep[e.g.,][]{fk01}.

The excess of transient X-ray sources within 1 pc of \sgrastar\ 
brings to mind the fact that, 
relative to their numbers in the rest of the Galaxy, LMXBs
are also a factor of $\sim 100$ over-abundant in globular clusters
\citep{katz75,cla75}. In the cores of globular clusters, three-body 
interactions produce this excess of LMXBs 
\citep[e.g., Rasio \etal\ 2000;][]{pool03}. 
We propose that the same process occurs near Sgr A$^*$, an idea
described fully in an upcoming paper by E. Pfahl \& A. Loeb.  Below is
an estimate of the number of dynamically-formed LMXBs in the
central parsec.

There may be $\ga$$10^4$ neutron stars and stellar-mass black holes
within 1\,pc of Sgr A$^*$, most driven there by dynamical friction
(Morris 1993; Miralda-Escud{\' e} \& Gould 2000).  We suppose that the
number densities of both neutron stars and black holes in the central
parsec are $n_x(r) = 10^3\pc^{-3}\,(r/{\rm pc})^{-2}$, roughly
$\simeq$1\% of the total stellar density (Genzel et al. 2003).  Some
of these compact objects may acquire a stellar companion following a
binary-single exchange encounter.  Binaries certainly populate the
central parsec, although their statistics are unknown.  We speculate
that $\sim$10\% of the stars are in binaries for $r > 0.1\pc$, and
assume a binary density of $n_b(r) = 10^4\pc^{-3}\,(r/{\rm pc})^{-2}$.  In
the central 0.1\,pc, stellar speeds of $\ga$$200 {\rm km~s}^{-1}$ 
make ionization
likely within $\sim$1\,Gyr for systems with initial semimajor axes of
$a \ga 0.1\au$.  We further assume for simplicity that the average
binary semimajor axis and systemic mass, $\langle a \rangle$ and
$\langle M_b \rangle$, are independent of $r$.  The issues of
dynamical evolution and survival of binaries will be more fully
discussed in E.\ Pfahl \& A.\ Loeb (in prep.).

For a given compact object of mass $M_x$ at radius $r < 1\pc$, the
exchange rate is $\gamma \sim n_b\Sigma \sigma$, where $\Sigma \simeq
2\pi \langle a \rangle G(\langle M_b \rangle + M_x)/\sigma^2$ is the
characteristic exchange cross section, $\sigma(r) = (GM_*/3r)^{1/2}$
is the one-dimensional velocity dispersion, and $M_* \simeq 3.7\times
10^6\emsun$ is the mass of Sgr A$^*$.  Numerically we find
\begin{equation}
\gamma \sim 10^{-12}\yr^{-1}
\left(\frac{\langle a \rangle}{0.1\au}\right)
\left(\frac{\langle M_b \rangle + M_x}{M_\odot}\right)
\left(\frac{r}{{\rm pc}}\right)^{-3/2}~.
\end{equation}
The rate density in the central parsec is $n_x \gamma \propto
r^{-7/2}$, while $n_x\gamma \propto r^{-4}$ in the roughly isothermal
region at $r \ga 2$--3\,pc.  This profile is much steeper than the
stellar density, which could partly explain the
overabundance of transients near Sgr A$^*$.  Volume integration of
$n_x \gamma$ over $r = 0.1$--1\,pc yields
\begin{equation}
\Gamma \sim 10^{-7}\yr^{-1} 
\left(\frac{\langle a \rangle}{0.1\au}\right)
\left(\frac{\langle M_b \rangle + M_x}{M_\odot}\right)~.
\end{equation}
For black holes, we expect $\langle M_b \rangle + M_x \simeq 10\emsun$
and $\Gamma_{\rm BH} \sim 10^{-6}\yr^{-1}$, while for neutron stars
$\Gamma_{\rm NS} \sim \text{few}\times 10^{-7}\yr^{-1}$.  Within the
central parsec, the dynamical friction time scale is $\sim$$10\,{\rm
Gyr}\,(M_x/M_\odot)^{-1}(r/{\rm pc})^{1/2}$.  This dynamical friction 
should limit the lifetime of an LMXB to $\sim$1\,Gyr.  Therefore,
perhaps 100--1000 neutron stars and black holes in the central parsec
have stellar companions.  Nearly all of these should be
luminous X-ray sources at some stage, in many cases driven by the
expansion of the star as it leaves the main sequence.  Thus, if $\la 1$\% 
of dynamically-formed LMXBs are in outburst each year, they 
can easily account for the overabundance of transients
in the central parsec.

In addition, since a concentration of young stars is observed within
the central parsec \citep[e.g.,][]{kra95}, some of the X-ray 
transients within 1 pc of \sgrastar\ could be HMXBs that formed recently.
The observed young, emission-line stars are thought to be the 
brightest few percent of $\sim 10^4$ stars that formed 
only $\approx 3-7$~Myr ago \citep{kra95}. However, the relative lack 
of older, K and M supergiants suggests that 
star formation was not sustained on time scales of tens of Myr. Since
this is the lifetime of a HMXB, any excess of such systems should be 
$\la 7$~Myr old. 

Predicting the numbers of HMXBs that have been produced in the 
central parsec is difficult, because the strong
tidal forces produced by \sgrastar\ makes it unlikely
that stars can form through the same mechanisms as in the Galactic disk
\citep{mor93}.  Instead, it has been suggested that the young 
stars in the central parsec were formed either {\it in situ} either after the 
collision of dense molecular clouds or in a massive disk around 
\sgrastar\ \citep[see, e.g.,][]{gen03,lb03,ml04}, or as a cluster at larger 
radii that settled dynamically into the central parsec 
\citep[e.g.,][]{ger01, pmg03, km03}. A crude estimate can be obtained
by assuming that HMXBs form with a comparable 
efficiency as in the Galactic disk, in which case $\sim 10$\% of 
binaries that initally contain a massive star should evolve into 
wind-accreting X-ray binaries (e.g., Pfahl, Rappaport, \& 
Podsiadlowski 2002)\nocite{prp02}. However, in $7$~Myr only stars more 
massive than $\simeq 30$~\msun\ should 
collapse to form compact objects. If we assume 
an initial mass function $dN/dM \propto M^{-2}$ and a 
minimum mass of 1~\msun\ \citep{kra95}, 
$\sim 3$\% of the $10^4$ stars will have $M > 30 M_{\odot}$. Therefore, 
$\sim 30$ HMXBs could have formed in the past 7~Myr. If these HMXBs are 
$\ga 10$ times more active than dynamically-formed
LMXBs, they could contribute to the transient population in the central parsec.

If the young stars in the central parsec were carried there by an in-falling
cluster, then a few HMXBs could have been formed through binary-single
interactions in the cluster cores. From eq.~(1) above, 
in a cluster with $n \sim 10^6$ pc$^{-3}$ \citep[e.g.,][]{km03}, 
a standard binary fraction of 0.5, and 
$\sigma \sim 10$ km s$^{-1}$, the interaction rate between a black hole and
binaries with $\langle a \rangle \sim 0.1$~AU will 
be $\gamma \sim 10^{-10}$ yr$^{-1}$. 
Assuming that there are initially $\sim 10^5$ stars in the cluster and 
$\sim 3$\% of them are black holes, then, in 7~Myr, $\sim$10 black 
hole binaries should form via three-body interactions. Thus,
HMXBs that were dynamically formed in an in-falling cluster 
could contribute to the number of transients in the central parsec.

Further progress can be made using several approaches. If variable infrared
or radio counterparts to these X-ray transients can be identified, it 
would enable us to determine whether the observed transients 
near \sgrastar\ are LMXBs or HMXBs. \chandra\ observations of 
similar star-forming environments, such as young, dense clusters, could 
establish how many HMXBs are formed along with the young stars. Finally, 
theoretical population synthesis studies could determine how efficiently 
HMXBs (not to mention their progenitor stars) can form in the extreme 
environment near \sgrastar.

\acknowledgments
We thank A. Loeb and G. Bellanger for helpful discussions, 
and H. Tananbaum for providing a discretionary \chandra\ observation 
to monitor one of the transients. MPM was supported by NASA and the Hubble 
Fellowship program through grant number HST-HF-01164.01-A, and EP 
by NASA and the Chandra 
Postdoctoral Fellowship program through grant number PF2-30024. WNB was 
supported by NSF CAREER Award 9983783.

\end{document}